\documentclass[12pt]{iopart}

\usepackage{iopams}
\usepackage{bbold}
\usepackage{amsopn}
\usepackage{calrsfs}
\usepackage{graphicx}
\usepackage{sidecap}
\usepackage{xcolor}

\definecolor{mycyan}{rgb}{0.,0.,0.5}

\usepackage[colorlinks]{hyperref}
\hypersetup{
	colorlinks   = true,
	citecolor    = mycyan,
	linkcolor    = mycyan,
	urlcolor     = magenta
}

\DeclareMathAlphabet{\pazocal}{OMS}{zplm}{m}{n}

\newcounter{numrel}

\newcommand{\ket}[1]{\left\vert#1\right\rangle}
\newcommand{\bra}[1]{\left\langle#1\right\vert}
\newcommand{\trace}[1]{\mbox{tr}\,\{#1\}}
\newcommand*\diff{\mathop{}\!\mathrm{d}}

\begin{document}
	
\title[Energy-efficient quantum frequency estimation]{Energy-efficient quantum frequency estimation}
	
\author{Pietro Liuzzo-Scorpo$^1$, Luis A. Correa$^1$, Felix A. Pollock$^2$, Agnieszka G\'orecka$^2$, Kavan Modi$^2$, and Gerardo Adesso$^1$}
\address{$^1$School of Mathematical Sciences and Centre for the Mathematics and Theoretical Physics of Quantum Non-Equilibrium Systems, University of Nottingham, University Park, Nottingham NG7 2RD, United Kingdom}
\vspace{5pt}
\address{$^2$School of Physics and Astronomy, Monash University, Clayton, Victoria 3800, Australia}
\ead{Luis.Correa@nottingham.ac.uk}
	
\vspace{10pt}
\begin{indented}
	\item[]October 2017
\end{indented}
	
\begin{abstract}
The problem of estimating the frequency of a two-level atom in a noisy environment is studied. Our interest is to minimise \textit{both} the energetic cost of the protocol and the statistical uncertainty of the estimate. In particular, we prepare a probe in a `GHZ-diagonal' state by means of a sequence of qubit gates applied on an ensemble of $ n $ atoms in thermal equilibrium. Noise is introduced via a phenomenological time-nonlocal quantum master equation, which gives rise to a phase-covariant dissipative dynamics. After an interval of free evolution, the $ n $-atom probe is globally measured at an interrogation time chosen to minimise the error bars of the final estimate. We model explicitly a measurement scheme which becomes optimal in a suitable parameter range, and are thus able to calculate the \textit{total} energetic expenditure of the protocol. Interestingly, we observe that scaling up our multipartite entangled probes offers no precision enhancement when the total available energy $ \pazocal{E} $ is limited. This is at stark contrast with standard frequency estimation, where larger probes---more sensitive but also more `expensive' to prepare---are always preferred. Replacing $ \pazocal{E} $ by the \textit{resource} that places the most stringent limitation on each specific experimental setup, would thus help to formulate more realistic metrological prescriptions. 	
\end{abstract}
	
\vspace{2pc}
	
\section{Introduction}\label{sec:intro}	

While (classical) metrology is concerned with producing the most accurate estimate of some relevant parameter, \textit{quantum} metrology is aimed at exploiting genuinely quantum traits to go beyond classical metrological limits \cite{giovannetti2004quantum, PhysRevLett.96.010401, giovannetti2011advances}. Classically, there would be no difference between running some estimation protocol sequentially $ N $ times on one probe, and running the same protocol simultaneously on $ n $ (uncorrelated) copies of that probe for $ M = N/n $ rounds. Quantum-mechanically, however, such $ n $-partite probe can be prepared in an entangled state, so that its \textit{estimation efficiency} grows \textit{super-extensively}.\footnote{Further improvements may follow from setting up interactions within the probe \cite{luis2004nonlinear, boixoext, choi2008bose, napolitano2011interaction, PhysRevA.94.042121, PhysRevLett.119.010403}, although such a scenario will not be considered in this paper.} Here `super-extensive' stands for faster-than-linear in the probe size, and the `estimation efficiency' is proportional to the inverse of the mean squared error.

More precisely, under rather weak conditions, the statistical uncertainty of the estimate of some parameter $ y = \bar{y} \pm \delta y $ may be tightly lower-bounded as $ \delta y \geq 1/\sqrt{M \mathcal{F}_y(\pmb O)} $ \cite{cramer1999mathematical, PhysRevLett.72.3439}, where $ \mathcal{F}_y(\pmb O) $ denotes the Fisher information of a sufficiently large number $ M $ of measurements of the observable $ \pmb O $ on the $ n $-partite probe. Importantly---although often disregarded---the length $ M $ of the dataset used to build the estimate will always be capped by the limited availability of some essential \textit{resource} $ \pazocal{R} $; that is, if $ r $ is the amount of resource consumed per round, $ M = \pazocal{R}/r $ and hence, $ \delta y \geq 1/\sqrt{\pazocal{R}\,\eta_{\pazocal{R}}} $, were $ \eta_{\pazocal{R}} \equiv \mathcal{F}_y(\pmb O)/r $ is the estimation efficiency. A scaling such as $ \eta_{\pazocal{R}}\sim n^c $, with $ c > 1, $ would be the hallmark of quantum-enhanced sensing.

Although the unavoidable effects of environmental noise often cancel out any quantum advantage \cite{PhysRevLett.79.3865,escher2011noisy,demkowicz2012elusive,PhysRevA.94.042101,Sekatskiquantum}, a super-extensive growth of the efficiency may still be attained under time-inhomogeneous phase-covariant noise \cite{PhysRevA.84.012103,PhysRevLett.109.233601,PhysRevA.92.010102,smirne2015ultimate}, and even more generic Ohmic dissipation \cite{haase2017fundamental}, noise with a particular geometry \cite{PhysRevLett.111.120401,PhysRevX.5.031010}, or setups involving quantum error correction \cite{PhysRevLett.112.150802,PhysRevLett.112.080801,PhysRevLett.116.230502}.

For instance, when it comes to frequency estimation, the total running time $ \pazocal{T} $ is usually regarded as the resource to be optimally partitioned \cite{PhysRevLett.79.3865}. Note that, even if features such as the amount of entanglement, coherence \cite{PhysRevA.88.042109}, or squeezing \cite{aasi2013enhanced} in the initial state of the probe, or the internal interaction range among its constituents \cite{luis2004nonlinear,boixoext,choi2008bose,napolitano2011interaction,PhysRevLett.119.010403} could all be regarded as legitimate \textit{metrological resources}, these do not fit in our framework. That is, even if, e.g., the amount of entanglement in the preparation of an $ n $-partite probe was severely limited in practice, this would not cap the number of rounds $ M $ of the estimation protocol---a fresh copy of the \textit{same} entangled state would be supplied at the start of every iteration until either time, the overall number of probe constituents, or the available energy have been fully consumed.

In our case, we shall look precisely at the total energy consumed $ \pazocal{E} $, and show that the notion of \textit{optimality} that follows from the maximisation of an energy efficiency differs fundamentally from the one based solely on the portioning of the available time. In particular, while the maximisation of a time efficiency encourages the use of multipartite entangled probes with $ n $ as large as possible, energetic considerations advice against it---the high costs associated with the creation and manipulation of large multipartite correlated states does not pay off from the metrological viewpoint. In this way, we put into qualitative terms the intuitive notion that multi-particle entanglement-enabled metrology may not always be practical \cite{Braunreview}.

\begin{figure*}[t!]
	\includegraphics[width=\linewidth]{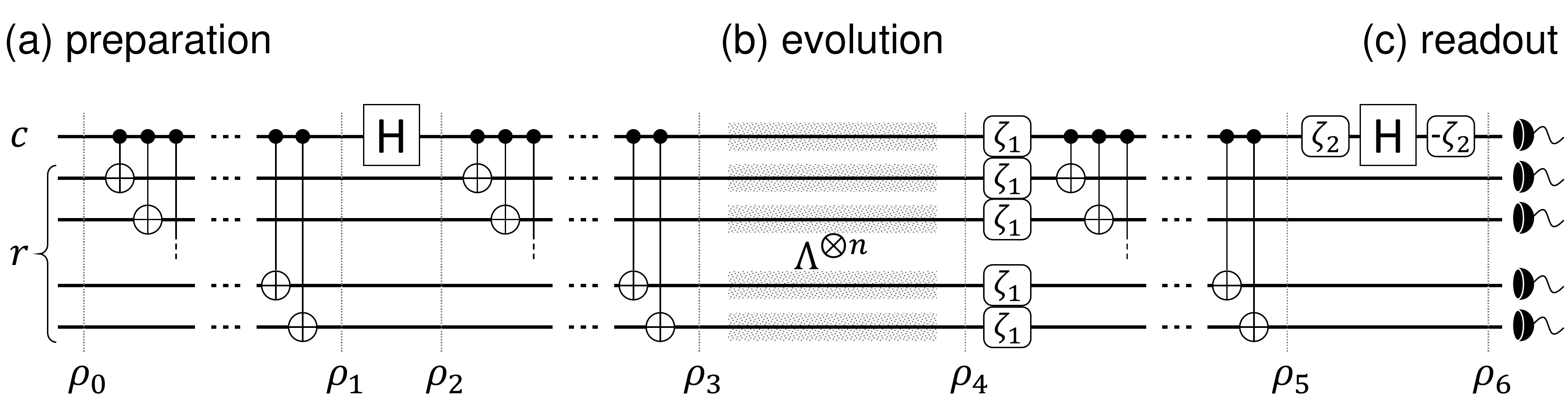}
	\caption{Circuit representation of the {\sf (a)} preparation, {\sf (b)} free evolution, and {\sf (c)} readout stages of our estimation protocol, as discussed in the main text. {\sf (a)} A probe system composed of $1$ control ($c$) qubit and $n-1$ register ($r$) qubits, initially in a thermal state $\pmb\rho_0$, is prepared into a GHZ-diagonal state $\pmb\rho_3$ by a sequence of \texttt{CNOT}, Hadamard [{\sf H}], and \texttt{CNOT} gates. {\sf (b)} The system is left to evolve freely for a time $t$ under a noisy environment according to a master equation with a memory kernel; this amounts to the action of the phase-covariant channel $\Lambda$, which imprints a phase $\phi = \omega t$ on the qubits while inducing dissipation effects, overall transforming the state of the system into $\pmb\rho_4$. {\sf (c)} A pre-measurement sequence of qubit rotations, \texttt{CNOT} gates, and a rotated Hadamard on the control qubit is applied, leading to the state $\pmb\rho_6$; each rounded rectangle $(\zeta)$ indicates a single-qubit rotation by an angle $\zeta$, described by the unitary $e^{-i \zeta \pmb\sigma_z/2}$. The system is finally measured in the energy basis to estimate the frequency $\omega$ with optimal efficiency.}
	\label{fig0}
\end{figure*}

In particular, as illustrated in figure~\ref{fig0}, we consider an ensemble of $ n $ initially thermal two-level atoms that are brought, through a sequence of qubit gates, into a sensitive GHZ-diagonal state \cite{modi2011mixed} (cf. section \ref{sec:preparation}). Such entangled probe is left to evolve freely under the action of time-non-local covariant noise. Specifically, we resort to a phenomenological quantum master equation \cite{PhysRevA.73.012111,PhysRevA.75.062103,PhysRevA.81.062120} which explicitly accounts for memory effects and gives rise to a non-divisible dissipative dynamics \cite{PhysRevA.81.062120} (see section \ref{sec:noise_model} for full details). We then devise a measurement protocol consisting of a sequence of qubit gates followed by an energy measurement (cf. section \ref{sec:readout}). We further provide the specific measurement setting for which this scheme becomes optimal for frequency estimation in a suitable parameter range (cf. section \ref{sec:qfi}). By looking at the changes in the average energy of the probe during the preparation and measurement stages, we explicitly obtain the total energetic cost per round. We find that adjusting the free evolution time so as to maximise the time efficiency of the protocol does lead to a super-extensive scaling in the probe size; specifically $ n^{3/2} $ or `Zeno scaling' \cite{PhysRevLett.109.233601,PhysRevA.92.010102}. In contrast, the energy efficiency of the very same probe, decays monotonically with $ n $, even when the time is chosen to maximise it (see section \ref{sec:results}).

Interestingly, note that the observed super-extensive growth of the time efficiency is attained while starting from thermal qubits that are prepared into a GHZ-diagonal state. In an accompanying article \cite{monashpaper} the same super-extensive growth of the time efficiency is found for an arbitrary set of qubits prepared in a GHZ-diagonal state for frequency estimation in a noisy environment. The GHZ-diagonal state had been conjectured to be optimal for phase estimation with mixed probes in the absence of noise \cite{modi2011mixed}. Here, we show that they lead to optimal scaling \textit{even in a noisy scenario}. We also observe that, in our setting, memoryless `Markovian' dissipative dynamics generally produces less efficient estimates, thus suggesting that memory effects might be beneficial for the energy efficiency of parameter estimation (cf. section \ref{sec:results}).

\section{Methods}

\subsection{Probe initialisation}
\label{sec:preparation}
	
The system of interest is an ensemble of $ n $ non-interacting two-level atoms thermalised at temperature $ T $, whose frequency $ \omega $ needs to be estimated. For simplicity of notation we shall set $ \hbar $ and the Boltzmann constant $ k_B $ to $ 1 $ in all what follows. Each atom has a Hamiltonian $ \pmb{h} = \frac{\omega}{2} \pmb{\sigma}_z $ and is initially in the state
\begin{equation}
\pmb\varrho = \frac12
\left(\begin{array}{cc}
1-\epsilon & 0 \\
0 & 1+\epsilon
\end{array}
\right),
\label{eq:initial_state}
\end{equation}
where the polarization bias $ \epsilon = \tanh{\big(\frac{\omega}{2T}\big)} $ so that $  \pmb{\varrho} \propto \exp{(- \pmb{h}/T)} $, and $  \pmb{\sigma}_z $ denotes the $ z $ Pauli matrix. The global Hamiltonian is $ \pmb{H} = \frac{\omega}{2} \pmb{J}_z $, where $ \pmb{J}_z = \pmb{\sigma}_z\otimes\mathbb{1}^{\otimes n-1} + \mathbb{1}\otimes \pmb{\sigma}_z\otimes\mathbb{1}^{\otimes n-2} + \cdots + \mathbb{1}^{\otimes n-1}\otimes \pmb{\sigma}_z $ and the total initial state is simply
\begin{equation}
\pmb\rho_0 = \pmb{\varrho}^{\otimes n} \equiv \pmb{\varrho}_c\otimes \pmb{\varrho}_r^{\otimes n-1} = \frac12
\left( \begin{array}{cc}
(1-\epsilon) \pmb{\varrho}^{\otimes n-1} & 0 \\
0 & (1+\epsilon) \pmb{\varrho}^{\otimes n-1}
\end{array} \right),
\label{eq:sigma_0}
\end{equation}
where we have labelled the first atom as $ c $ for `control qubit' while the rest are tagged $ r $, for `register'. 

We shall prepare our $ n $-atom probe in a GHZ-diagonal state by means of a \texttt{CNOT} transformation, followed by a Hadamard gate and a further \texttt{CNOT} [see figure~\ref{fig0}(a)] \cite{modi2011mixed}. That is, we first apply the unitary $ \ket{0}_c\bra{0}_c\otimes\mathbb{1}^{\otimes n-1} + \ket{1}_c\bra{1}_c\otimes \pmb{\sigma}_x^{\otimes n-1}$ on $ \pmb\rho_0 $. Introducing the denotation $ \pmb{\bar{A}} \equiv \pmb\sigma_x \pmb A \pmb\sigma_x $, this yields
\begin{equation}
\pmb{\rho}_1 = \frac12\left( \begin{array}{cc}
(1-\epsilon) \pmb{\varrho}^{\otimes n-1} & 0 \\
0 & (1+\epsilon)\pmb{\bar{\varrho}}^{\otimes n-1}
\end{array}
\right).
\label{eq: rho_1}
\end{equation}
Then, the Hadamard transformation $ \pmb U_H \equiv \frac{1}{\sqrt{2}}(\pmb\sigma_x + \pmb\sigma_z)\otimes\mathbb{1}^{n-1} $ acts solely on the control qubit:
\begin{equation}
\pmb\rho_2 = \frac{1-\epsilon}{4}\left(\begin{array}{cc}
\pmb\varrho^{\otimes n-1} & \pmb\varrho^{\otimes n-1} \\
\pmb\varrho^{\otimes n-1} & \pmb\varrho^{\otimes n-1}
\end{array}
\right) +
\frac{1+\epsilon}{4}\left(\begin{array}{cc}
\pmb{\bar{\varrho}}^{\otimes n-1} & -\pmb{\bar{\varrho}}^{\otimes n-1} \\
-\pmb{\bar{\varrho}}^{\otimes n-1} & \pmb{\bar{\varrho}}^{\otimes n-1}
\end{array}
\right),
\label{eq: rho_2}
\end{equation}
and finally, the second \texttt{CNOT} transformation leads to
\begin{equation}
\pmb\rho_3 = \frac{1-\epsilon}{4}\left(\begin{array}{cc}
\pmb\varrho^{\otimes n-1} & (\pmb\varrho\pmb\sigma_x)^{\otimes n-1} \\
\mbox{h.c.} & \pmb{\bar{\varrho}}^{\otimes n-1}
\end{array}
\right) +
\frac{1+\epsilon}{4}\left(\begin{array}{cc}
\pmb{\bar{\varrho}}^{\otimes n-1} & -(\pmb\sigma_x\pmb\varrho)^{\otimes n-1} \\
\mbox{h.c.} & \pmb\varrho^{\otimes n-1}
\end{array}
\right),
\label{eq: rho_3}
\end{equation}
where the missing elements are just Hermitian conjugates of the opposite corners of each matrix. The resulting state will subsequently undergo dissipative evolution (cf. section~\ref{sec:noise_model}) before being interrogated.

As we will see in section \ref{sec:noise_model}, our model of dissipation gives rise to phase-covariant dynamics. It is known that the mean squared error of frequency estimated with this type of noise can be tightly lower-bounded \textit{below} the standard quantum limit \cite{smirne2015ultimate,PhysRevA.92.010102}. It was further shown that this bound is asymptotically saturable by using (pure) GHZ input states. On the other hand, (mixed) GHZ-diagonal states such as $ \pmb\rho_3 $ were found to perform well---and conjectured to be optimal---in \textit{noiseless} phase estimation with \textit{mixed} probes \cite{modi2011mixed}. In section \ref{sec:results} we will illustrate that the optimal `Zeno scaling', introduced in references \cite{PhysRevLett.109.233601, PhysRevA.92.010102}, can also be attained with such GHZ-diagonal states. 

Even though in the present paper we will limit ourselves to GHZ-diagonal preparations, it seems interesting to compare the size scaling of the metrological performance of \textit{different} preparations. One would certainly find that some preparations may allow for a more energy-efficient estimation than others \textit{at fixed probe size}. Unfortunately, as we will see below, our calculations rely heavily on the simple analytical structure of GHZ-diagonal states undergoing phase-covariant dissipation. This makes it difficult to extrapolate our results to other initial states. 

Finally, note that the energetic cost of this initialisation stage $ \pazocal{E}_{init} = \trace{\pmb H(\pmb\rho_3-\pmb\rho_0)}$ is linear in the probe size and evaluates to
\begin{equation}
\pazocal{E}_{init} = \frac12 \omega n \epsilon.
\label{eq:energy_preparation}
\end{equation}

At this point, one may wonder why do we not cool down probes to the ground state before starting the estimation protocol so as to work with pure rather than mixed states. This could certainly be done (e.g. by coherent feedback cooling), so long as the corresponding energy cost $ \pazocal{E}_{cool} $ is added to the total energetic bookkeeping---just like (\ref{eq:energy_preparation}), $ \pazocal{E}_{cool} $ would scale linearly in $ n $. Such cooling stage is anyway not essential, and we will keep it out of the picture in what follows, thus avoiding to model it explicitly. 

\subsection{Free evolution}\label{sec:noise_model}

\subsubsection{Phenomenological master equation---\kern -0.8em}

In order to account for the environmental effects in our probe, we will assume that each atom evolves according to a \textit{time-nonlocal} master equation [see figure~\ref{fig0}(b)] with a phenomenological exponentially-decaying memory kernel \cite{PhysRevA.73.012111}. The reason for this choice is that the resulting dissipative dynamics is phase-covariant, as opposed to the one following from a more canonical setting, such as the spin-boson model \cite{breuer2002theory,haase2017fundamental}. This will eventually allow us to establish a connection with known results in the literature \cite{smirne2015ultimate}. Moreover, due to its simplicity, the model considered here can be solved exactly.

Specifically, we shall think of a generic scenario in which a two-level atom with Hamiltonian $ \pmb h $ interacts with a bath ($ \pmb H_B $) through the interaction term $ \pmb H_{int} $. In the interaction picture with respect to the free Hamiltonian $\pmb H_0 = \pmb h + \pmb H_B $ (indicated with subindex $ I $ in what follows), our phenomenological equation would read
\begin{equation}
\frac{\diff \pmb\varrho_I}{\diff t} = \int_0^t \diff s f(t-s) \pazocal{L}\pmb\varrho_I(s),
\label{eq:phenomenological_master}
\end{equation}
with $ f(t) \equiv \lambda e^{-\lambda \vert t\vert } $ and where $ \pazocal{L} $ denotes the Gorini-Kossakowski-Lindblad-Sudarshan (Markovian) generator \cite{lindblad1976generators,gorini1976completely}	
\begin{eqnarray}
\pazocal{L}\pmb\varrho_I \equiv &\Gamma_{\omega} \left(\pmb\sigma_-\pmb\varrho_I\,\pmb\sigma_+ -\frac12 \{\pmb\sigma_+\pmb\sigma_-,\pmb\varrho_I\}_+\right) \nonumber \\
&+ \Gamma_{-\omega} \left(\pmb\sigma_+\pmb\varrho_I\,\pmb\sigma_- -\frac12 \{\pmb\sigma_-\pmb\sigma_+,\pmb\varrho_I\}_+\right).
\label{eq:linbladian}
\end{eqnarray}
Here $ \{\cdot,\cdot\}_+ $ stands for anti-commutator, and the decay rates are $ \Gamma_{\omega} \equiv \gamma_0 [1 + (e^{\omega/T}-1)^{-1}] $ and $ \Gamma_{-\omega} = e^{-\omega/T}\,\Gamma_{\omega} $. Equation (\ref{eq:phenomenological_master}) comes with the advantage of explicitly introducing memory effects into the dynamics. Note, however, that one must be careful when dealing with master equations that lack a microscopic derivation \cite{levy2014local,PhysRevA.64.033808,stockburger2016thermodynamic} as they often lead to unphysical results. In particular, equation~(\ref{eq:phenomenological_master}) breaks positivity iff $ \frac{\gamma_0}{\lambda\epsilon}\geq \frac14 $ \cite{PhysRevA.75.062103}. Importantly, the thermal sate $ \pmb\varrho $ is the stationary point equation (\ref{eq:phenomenological_master}), which is, in turn, consistent with our choice of initial state in section \ref{sec:preparation}.

At this point, one may still wonder why not to choose an arguably more realistic non-covariant noise model derived from first principles, as in reference \cite{haase2017fundamental}. It must be noted that---unlike in \cite{haase2017fundamental}---we need to know the explicit form of the time-evolved state for arbitrarily large probes. This is a prerequisite for gauging the energy cost of the measurement stage, and, eventually, assessing the asymptotic scaling of the overall estimation efficiency. A noise model lacking the ``niceties'' of covariant channels not only does compromise our ability to analytically evolve the state of the probe, but is also likely to render our proposed measurement scheme sub-optimal. On the plus side, however, covariant dissipation follows quite naturally from generic noise models whenever the \textit{ubiquitous} rotating-wave approximation is well justified \cite{breuer2002theory,haase2017fundamental}. Furthermore, as it can be seen by comparing \cite{smirne2015ultimate} with \cite{monashpaper} and our results below, the details of the specific covariant dissipation model do not seem to affect the qualitative asymptotic features of the estimation protocol.

\subsubsection{Connection to the damped Jaynes-Cummings model---\kern -0.8em}

The seemingly arbitrary choice of memory kernel in equation~(\ref{eq:phenomenological_master}) may be justified by considering the damped Jaynes-Cummings model on resonance; that is, a two-level atom in an empty and leaky cavity. This setup can be effectively described by the Hamiltonian
\begin{equation}
\pmb H_{JC} = \frac{\omega}{2}\pmb\sigma_z + (\pmb\sigma_+ \pmb B + \pmb\sigma_- \pmb B^\dagger) + \sum_{\mu} \omega_\mu \pmb b_\mu^\dagger \pmb b_\mu,
\label{eq:Jaynes-Cummings_ham}
\end{equation}
where $ \pmb B \equiv \sum_{\mu} g_\mu (\pmb b_\mu + \pmb b_\mu^\dagger) $ and the system-bath coupling constants $ g_\mu $ make up the Lorentzian spectral density $ J(\omega) = \sum_{\mu} g_\mu^2\,\delta(\omega-\omega_\mu) = \frac{1}{2\pi}\frac{\gamma_0 \lambda^2}{(\omega-\omega)^2 + \lambda^2} $ \cite{breuer2002theory,PhysRevA.73.012111}.

Assuming \textit{weak coupling}, the use of a second-order Nakajima-Zwanzig master equation \cite{nakajima1958quantum,zwanzig1960ensemble,breuer2002theory} is justified. This reads
\begin{equation}
\frac{\diff \pmb\varrho_I }{\diff t} = -\int_0^t \diff s \tr_B{\left[\pmb H_{JC}(t),\left[\pmb H_{JC}(s),\pmb\varrho_I(s)\otimes\pmb\varrho_B\right]\right]},
\label{eq:nakajima_zwanzig}
\end{equation}-
where the interaction picture Hamiltonian is $ \pmb H_{JC}(t) = \pmb\sigma_+(t)\pmb B(t)+\pmb\sigma_-(t)\pmb B^\dagger(t) $, with $ \pmb\sigma_\pm(t) = \pmb\sigma_\pm e^{\pm i\omega t} $ and $ \pmb B(t) = \sum_{\mu} g_\mu (\pmb b_\mu e^{-i\omega_\mu t} + \pmb b_\mu^\dagger e^{i\omega_\mu t}) $. The state of the environment and the trace over its degrees of freedom are denoted by $ \pmb\varrho_B $ and $ \tr_B $, respectively.

Combining equations (\ref{eq:Jaynes-Cummings_ham}) and (\ref{eq:nakajima_zwanzig}) one arrives to a master equation with the same structure as (\ref{eq:phenomenological_master}) at zero temperature \cite{breuer2002theory}, in which the bath correlation function $ \langle \pmb B(t)\pmb B^\dagger(s)\rangle = \int \diff \omega' J(\omega') e^{i(\omega-\omega')(t-s)} = \frac{\gamma_0\lambda}{2} e^{-\lambda t} $ plays the role of the memory kernel. In spite of this remark, we emphasise that (\ref{eq:phenomenological_master}) remains a purely \textit{phenomenological} equation, as the decay rates $ \Gamma_\omega $ are evaluated at arbitrary temperature $ T $.

\subsubsection{Dissipative dynamics as a phase-covariant channel---\kern -0.8em}

Alternatively, (\ref{eq:phenomenological_master}) can be brought into the Schr\"{o}dinger picture and cast in the equivalent \textit{time-local} form
\begin{eqnarray}
\frac{\diff \pmb\varrho}{\diff t} &= -i[\pmb h, \pmb \varrho] + \gamma_+(t) \left(\pmb\sigma_+\pmb\varrho\,\pmb\sigma_--\frac12\{\pmb\sigma_-\pmb\sigma_+,\pmb\varrho \}_+\right) \nonumber\\
&+ \gamma_-(t) \left(\pmb\sigma_-\pmb\varrho\,\pmb\sigma_+-\frac12\{\pmb\sigma_+\pmb\sigma_-,\pmb\varrho \}_+\right)
 + \gamma_z(t) \left(\pmb\sigma_z\pmb\varrho\,\pmb\sigma_z-\pmb\varrho \right).
\label{eq:master_equation_local}
\end{eqnarray}
For the sake of completeness, we include here the time-dependent decay rates $ \gamma_\pm(t) $ and $ \gamma_z(t) $, derived in reference~\cite{PhysRevA.81.062120}
\begin{eqnarray}
\gamma_\pm(t) &= -\frac12(1\mp\epsilon)\frac{\diff}{\diff t}\log{\xi_R(t)}\mbox{ and}\\
\gamma_z(t) &= \frac14\frac{\diff}{\diff t}\log{\frac{\xi_R(t)}{\xi_{R/2}^2(t)}},
\label{eq:rates_time_local}
\end{eqnarray}
where $ \xi_R(t) \equiv e^{-\lambda t/2}\left[ \frac{1}{\sqrt{1-4 R}}\sinh{\left(\frac{\lambda t}{2}\sqrt{1-4 R}\right)} + \cosh{\left(\frac{\lambda t}{2}\sqrt{1-4 R}\right)}\right] $ and $ R = \frac{\gamma_0}{\lambda \epsilon} $.	

As argued in \cite{smirne2015ultimate}, the dissipative dynamics following from equations such as (\ref{eq:master_equation_local}) can be cast a \textit{phase-covariant} qubit channel $ \pmb\varrho(t) = \Lambda(t)[\pmb\varrho(0)] $, i.e. a map such that $ \Lambda\circ\pazocal{U}_\varphi = \pazocal{U}_\varphi\circ\Lambda $, where $ \pazocal{U}_\varphi\,\pmb\varrho \equiv e^{-i \pmb h \varphi}\varrho e^{i \pmb h \varphi} $ and `$ \circ $' stands for channel composition. These maps can be parametrised as
\begin{equation}
\mathsf{\Lambda}(t) = \left( \begin{array}{cccc}
1 & 0 & 0 & 0 \\
0 & \eta_\perp(t)\cos{\omega t} & -\eta_\perp(t)\sin{\omega t} & 0 \\
0 & \eta_\perp(t)\sin{\omega t} & \eta_\perp(t)\cos{\omega t} & 0 \\
\kappa(t) & 0 & 0 & \eta_\parallel(t)
\end{array}\right),
\label{eq:matrix_channel}
\end{equation}
where the matrix $ \mathsf{\Lambda}(t) $ acts on $ \mathsf{v}(0) =  \big(1,\trace{\pmb\sigma_x\pmb\varrho(0)},\trace{\pmb\sigma_y\pmb\varrho(0)},\trace{\pmb\sigma_z\pmb\varrho(0)}\big) $ to yield $ \mathsf{v}(t) = \mathsf{\Lambda}(t)\mathsf{v}(0) $, so that $ \pmb\varrho(t) = \frac{1}{2} (\mathsf{v}_1(t) \mathbb{1} + \mathsf{v}_2(t) \pmb\sigma_x + \mathsf{v}_3(t)\pmb\sigma_y + \mathsf{v}_4(t)\pmb\sigma_z) $.

For the ensuing dynamics to be completely positive, one must have $ \eta_\parallel(t) \pm \kappa(t) \leq 1 $ and $ 1 + \eta_\parallel(t) \geq \sqrt{4\eta_\perp^2(t) + \kappa^2(t)} $. Additionally, since the map describes the action of the environment, it should asymptotically bring the two-level atom back to thermal equilibrium. This entails $ \kappa(\infty) = -\epsilon[1-\eta_z(\infty)] $.

Following \cite{smirne2015ultimate} one readily finds that equation~(\ref{eq:phenomenological_master}) corresponds to
\begin{eqnarray}
\eta_\alpha(t) &= \frac{e^{-t\lambda(1 + A_\alpha)/2}}{2 A_\alpha}\left[e^{t\lambda A_\alpha}(1 + A_\alpha)+ A_\alpha-1\right]\nonumber\mbox{ and}\\
\kappa(t) &= -\epsilon[1-\eta_\parallel(t)],
\label{eq:noise_parameters}
\end{eqnarray}
where $ \alpha\in\{\parallel,\perp \} $, $ A_\parallel = \sqrt{1-4 R} $, and $ A_\perp = \sqrt{1-2 R} $.

\subsubsection{State of the probe after the noisy evolution---\kern -0.8em} Having discussed the details of the noise model, let us explicitly write the time-evolved state $ \pmb\rho_4 \equiv \Lambda[\pmb\rho_3]^{\otimes n} $ after the action of the channel of equations (\ref{eq:matrix_channel}) and (\ref{eq:noise_parameters}). Its application to a generic qubit state yields
\begin{equation}
\Lambda\left[\left(\begin{array}{cc}
a & c \\
c^* & b
\end{array}\right)\right] =
\left(\begin{array}{cc}
a \alpha_1 + b \alpha_{-1} & c e^{-i\varphi}\eta_\perp \\
c^* e^{i\varphi}\eta_\perp & a \beta_1 + b \beta_{-1}
\end{array}\right),
\label{eq:channel_on_qubit_generic}
\end{equation}
with $ \alpha_s \equiv \frac12(1+s\eta_\parallel+\kappa) $, $ \beta_s \equiv \frac12(1-s\eta_\parallel-\kappa) $, and $ \varphi \equiv \omega t $. As a result
\begin{eqnarray}
&\pmb\rho_4 = \frac{1-\epsilon}{4}\left(\begin{array}{cc}
\alpha_1\Lambda[\pmb\varrho]^{\otimes n-1} + \alpha_{-1}\Lambda[\pmb{\bar{\varrho}}]^{\otimes n-1} & e^{-i\varphi}\eta_\perp\Lambda[\pmb\varrho\pmb\sigma_x]^{\otimes n-1} \\
\mbox{h.c.} & \beta_1\Lambda[\pmb\varrho]^{\otimes n-1} + \beta_{-1}\Lambda[\pmb{\bar{\varrho}}]^{\otimes n-1}
\end{array}\right) \nonumber\\
&+\frac{1+\epsilon}{4}\left(\begin{array}{cc}
\alpha_{-1}\Lambda[\pmb\varrho]^{\otimes n-1} + \alpha_{1}\Lambda[\pmb{\bar{\varrho}}]^{\otimes n-1} & -e^{-i\varphi}\eta_\perp\Lambda[\pmb\sigma_x\pmb\varrho]^{\otimes n-1} \\
\mbox{h.c.} & \beta_{-1}\Lambda[\pmb\varrho]^{\otimes n-1} + \beta_{1}\Lambda[\pmb{\bar{\varrho}}]^{\otimes n-1}
\end{array}\right),
\label{eq:rho_4}
\end{eqnarray}
where we have dropped the explicit time dependence from the noise parameters for brevity. We shall not attach any energetic cost to this stage of the estimation protocol as it corresponds to \textit{free} dissipative evolution.

\subsection{Probe readout}\label{sec:readout}

Before the probe is interrogated, it will need to undergo a \textit{pre-measurement} stage, consisting of sequence of three unitaries: First, each atom will be rotated by an angle $ \zeta_1 $ via $ \pazocal{U}^{\otimes n}_{\zeta_1} $. Then, a \texttt{CNOT} transformation and the generalised Hadamard gate
\begin{equation}
\pmb U_H(\zeta_2) = e^{-i\frac{\zeta_2}{2}\pmb{\sigma}_z} \pmb{U}_H e^{i\frac{\zeta_2}{2}\pmb{\sigma}_z} = \frac{1}{\sqrt{2}}\left(\begin{array}{cc}
1 & e^{-i\zeta_2} \\
e^{i\zeta_2} & -1
\end{array}\right)\otimes\mathbb{1}^{n-1},
\label{eq:generalised_hadamard}
\end{equation}
will be sequentially applied [see figure~\ref{fig0}(c)]. An energy measurement can then be performed on the probe in order to build the frequency estimate. As we shall argue in section \ref{sec:qfi} below, in the limit $ R \ll 1 $, the angles $ (\zeta_1,\zeta_2) $ may be chosen so that the statistical uncertainty of the resulting estimate is (nearly) minimal.

Let us thus obtain the probabilities associated with an energy measurement on the final state of the probe. The state after $ \pazocal{U}_{\zeta_1}^{\otimes n-1} $ and the \texttt{CNOT} transformation reads
\begin{eqnarray}
\pmb \rho_5 = \frac{1-\epsilon}{4}\left(\begin{array}{cc}
\alpha_1 \Lambda[\pmb \varrho]^{\otimes n-1} + \alpha_{-1}\Lambda[\pmb{\bar{\varrho}}]^{\otimes n-1} & e^{-i\phi} \eta_\perp(\Lambda[\pmb\varrho\pmb\sigma_x]\pmb\sigma_x)^{\otimes n-1} \\
\mbox{h.c.} & \beta_1 \overline{\Lambda[\pmb\varrho]}^{\otimes n-1} + \beta_{-1}\overline{\Lambda[\pmb{\bar{\varrho}}]}^{\otimes n-1}
\end{array}\right) \nonumber \\
+\frac{1+\epsilon}{4}\left(\begin{array}{cc}
\alpha_{-1} \Lambda[\pmb \varrho]^{\otimes n-1} + \alpha_{1}\Lambda[\pmb{\bar{\varrho}}]^{\otimes n-1} & -e^{-i\phi} \eta_\perp(\Lambda[\pmb\sigma_x\pmb\varrho]\pmb\sigma_x)^{\otimes n-1} \\
\mbox{h.c.} & \beta_{-1} \overline{\Lambda[\pmb\varrho]}^{\otimes n-1} + \beta_{1}\overline{\Lambda[\pmb{\bar{\varrho}}]}^{\otimes n-1}
\end{array}\right),
\label{eq:rho_5}
\end{eqnarray}
where $ \phi \equiv \omega t + \zeta_1 $, i.e. the action of $ \pazocal{U}^{\otimes n-1}_{\zeta_1} $ amounts to replacing $ \varphi\rightarrow\varphi + \zeta_1 $ in (\ref{eq:rho_4}).

It will be more convenient to cast $ \pmb\rho_5 $ in an alternative form. To that end, note that $ \Lambda[\pmb\varrho] = \alpha_{-\epsilon}\ket{0}\bra{0} + \beta_{-\epsilon} \ket{1}\bra{1} $, whereas $ \Lambda[\pmb\varrho]^{\otimes 2} = \alpha_{-\epsilon}^2\ket{00}\bra{11} + \alpha_{-\epsilon}\beta_{-\epsilon} (\ket{01}\bra{01} + \ket{10}\bra{10}) + \beta_{-\epsilon}^2\ket{11}\bra{11} $. Generalising to an arbitrary power $ l $ yields
\begin{equation}
\Lambda[\pmb\varrho]^{\otimes l} = \sum_{x=0}^{2^l-1}\alpha_{-\epsilon}^{h(\bar{x_l})}\beta_{-\epsilon}^{h(x_l)}\ket{x_l}\bra{x_l},
\label{eq:tensor_power_channel_on_thermal}
\end{equation}
where $ x_l $ stands for the $ l $-digit binary representation of $ x $ and $ h(x) $ denotes the number of non-zero digits in $ x_l $ (i.e. its Hamming weight). In turn, $ \bar{x}_l $ represents the bitwise negation of $ x_l $. Care must be taken not to confuse the scalar function $ h(\cdot) $ with the single-atom Hamiltonian $ \pmb h $, nor the bitwise negation $ \bar{x}_l $ with the map $ \pmb{\bar{\varrho}}=\pmb\sigma_x\pmb\varrho\pmb\sigma_x $.

Quantities such as $ \Lambda[\pmb{\bar{\varrho}}]^{\otimes l} $, $ \overline{\Lambda[\pmb\varrho]} $, and $ \overline{\Lambda[\pmb{\bar{\varrho}}]} $ follow from equation~(\ref{eq:tensor_power_channel_on_thermal}) by making the replacements $ -\epsilon \rightarrow \epsilon $, $ \alpha_{-\epsilon}\rightarrow\beta_{-\epsilon} $, and $ \alpha_{-\epsilon}\rightarrow\beta_{\epsilon} $, respectively, while $ \pmb\sigma_x^{\otimes l}\ket{\bar{x}_l} = \ket{x_l} $, and
\begin{equation}
\Lambda[\pmb\varrho\pmb\sigma_x]^{\otimes l} = \eta_\perp^l\sum_{x=0}^{2^l-1}\left(\frac{1-\epsilon}{2}\right)^{h(\bar{x}_l)}\left(\frac{1+\epsilon}{2}\right)^{h(x_l)}e^{-i\varphi[h(\bar{x}_l)-h(x_l)]}\ket{x_l}\bra{\bar{x}_l}.
\label{eq:auxiliary_rho_5}
\end{equation}

Putting together all the above and dropping the sub-indices $ l = n-1 $ in the interest of a lighter notation yields
\begin{equation}
\pmb\rho_5 = \sum_{x=0}^{2^{n-1}-1}\left(\begin{array}{cc}
a_x & e^{-i\phi f(x)}c_x \\
e^{i\phi f(x)}c_x & b_x
\end{array}\right)\otimes\ket{x}\bra{x},
\label{eq:rho_5_alt}
\end{equation}	
with the definitions
\begin{eqnarray}
a_x \equiv \frac12\left(\alpha_{-\epsilon}^{h(\bar{x})+1}\beta_{-\epsilon}^{h(x)} + \alpha_{\epsilon}^{h(\bar{x})+1}\beta_{\epsilon}^{h(x)}\right)\nonumber, \\
b_x \equiv \frac12\left(\alpha_{-\epsilon}^{h(x)}\beta_{-\epsilon}^{h(\bar{x})+1} + \alpha_{\epsilon}^{h(x)}\beta_{\epsilon}^{h(\bar{x})+1}\right)\nonumber, \\
c_x \equiv \frac{\eta_\perp^n}{2^{n+1}}\left[ (1-\epsilon)^{h(\bar{x})+1}(1+\epsilon)^{h(x)} - (1-\epsilon)^{h(x)}(1+\epsilon)^{h(\bar{x})+1} \right] \nonumber, \mbox{ and}\\
f(x) \equiv h(\bar{x}) - h(x) + 1.
\label{eq:abcf_parameters}
\end{eqnarray}
Similarly, the final state of the protocol [i.e. $ \pmb \rho_6 = \pmb U_H(\zeta_2)\,\pmb\rho_5\,\pmb U_H^\dagger(\zeta_2) $] is
\begin{equation}
\pmb\rho_6 = \sum_{x=0}^{2^{n-1}-1}\left(\begin{array}{cc}
\tilde{a}_x & e^{-i\zeta_2}\tilde{c}_x \\
e^{i\zeta_2}\tilde{c}_x^* & \tilde{b}_x
\end{array}\right)\otimes \ket{x}\bra{x},
\label{eq:rho_6}
\end{equation}
where
\begin{eqnarray}
\tilde{a}_x &\equiv \frac12[a_x + b_x + 2 c_x\cos{(\zeta_2-f(x)\phi)}] \nonumber,\\
\tilde{b}_x &\equiv \frac12 [a_x + b_x -2 c_x\cos{(\zeta_2-f(x)\phi)}] \nonumber,\mbox{ and}\\
\tilde{c}_x &\equiv \frac12 [a_x - b_x -2i c_x\sin{(\zeta_2-f(x)\phi)}].
\end{eqnarray} 	
Therefore, a measurement of $ \pmb\rho_6 $ in the energy basis $ \{\ket{0}\otimes\ket{x},\ket{1}\otimes\ket{x}\} $ has the following associated probabilities
\begin{eqnarray}
p_{0,h(x)} = \bra{0,x}\pmb\rho_6\ket{0,x} = \frac12[a_x + b_x + 2 c_x\cos{[\zeta_2-f(x)(\omega t + \zeta_1)]}] \nonumber\mbox{ and}\\
p_{1,h(x)} = \bra{1,x}\pmb\rho_6\ket{1,x} = \frac12[a_x + b_x - 2 c_x\cos{[\zeta_2-f(x)(\omega t + \zeta_1)]}],
\label{eq:probabilities}
\end{eqnarray}
where all eigenvectors with the same number of $ 1 $s [i.e. $ h(x) $] on the register yield the same probability. Equation (\ref{eq:probabilities}) will be used below to obtain a saturable lower bound on the mean squared error of the resulting frequency estimate.

We now look into the energetic cost of the pre-measurement stage  $ \pazocal{E}_{meas} = \pazocal{E}(\pmb\rho_6) - \pazocal{E}(\pmb\rho_4) $. Let us re-write the system Hamiltonian in the same notation as equations (\ref{eq:rho_5_alt}) and (\ref{eq:rho_6}). That is,
\begin{equation}
\pmb H = -\frac{\omega}{2}\sum_{x=0}^{n-1} [(h(x)-h(\bar{x})-1)\ket{0,x}\bra{0,x} + (h(x)-h(\bar{x})+1)\ket{1,x}\bra{1,x}].
\label{eq:alternative_hamiltonian}
\end{equation} 	
Hence, $ \pazocal{E}(\pmb\rho_4) \equiv \tr{\{\pmb H \pmb\rho_4\}} $ writes as
\begin{equation}
\pazocal{E}(\pmb\rho_4) = -\frac{\omega}{2}\sum_{x=0}^{2^{n-1}-1}[(h(x)-h(\bar{x})-1)a_x + (h(\bar{x})-h(x)+1)b_x] = \frac{\omega}{2}n\kappa,
\label{eq:energy_rho_4}
\end{equation}
whereas
\begin{eqnarray}
\pazocal{E}(\pmb\rho_6) = \tr{\{ \pmb H\pmb\rho_6 \}} = \sum_{x=0}^{2^{n-1}-1} [(h(x)-h(\bar{x})-1)\tilde{a}_x + (h(x)-h(\bar{x})+1)\tilde{b}_x] \nonumber \\ =\frac{\omega}{2}(n-1)(\epsilon^2\eta_\parallel^2+\kappa^2) + \omega\sum_{m=0}^{n-1} {{n-1}\choose{m}} c_m \cos{[\zeta_2-f_m(\omega t + \zeta_1)]},
\label{eq:energy_rho_6}
\end{eqnarray}
where the sub-indices $ m $ indicate the Hamming weight $ m = h(x) $ of the argument $ x $ of the corresponding coefficients, i.e. $ c_x $ and $ f_x $. At our optimal prescription $ (\zeta_1,\zeta_2) $ the pre-measurement energetic cost is always positive $ \pazocal{E}_{meas} > 0 $. 

Note that we are deliberately leaving the projective part of the measurement out of our energetic bookkeeping. In some setups such as nuclear magnetic resonance, this could be justified, as projective measurements are mimicked by suitable rotations followed by free decay. In other cases it may be necessary to supplement $ \pazocal{E}_{meas} $ with a `projection cost' $ \pazocal{E}_{proj} $. Similarly, depending on the specific projection model, the \textit{sharp} probabilities in equation (\ref{eq:probabilities}) might need to be modified---a `measurement apparatus' at some finite temperature would arguably introduce thermally distributed random bit flips during the readout, thus making the measurement \textit{noisy}. Neither the potential extra cost nor the errors in the interrogation would qualitatively affect our results.

While very general models of projective measurement schemes, and thermodynamic analyses thereof, may be found in the literature (see e.g. references \cite{sagawa2009minimal,erez2012thermodynamics,jacobs2012quantum,micadei2013thermodynamic,faist2015minimal,abdelkhalek2016fundamental,kammerlander2016coherence}, just to mention some), it is not our intention to make generic statements about the energy efficiency of frequency estimation. Instead, we settle for showing how looking at the energetic aspect of parameter estimation in a specific example can in fact change dramatically the usual notions of metrological optimality.	
	
\subsection{`Error bars' of the estimate} \label{sec:qfi}

\subsubsection{(Classical) Fisher information---\kern -0.8em}Recall from section \ref{sec:intro} that the mean squared error of a frequency estimate $ \omega = \bar{\omega} \pm \delta\omega $ constructed from a sufficiently large number of measurements $ M $ of some generic observable $ \pmb O $, can be tightly lower-bounded as $ \delta\omega\geq 1/\sqrt{M \mathcal{F}_\omega(\pmb O)} $ \cite{barndorff2000fisher}, where $ \mathcal{F}_\omega(\pmb O) $ stands for the (classical) Fisher information. In our case, $ \mathcal{F}_\omega(\pmb H) $ can be readily computed from the probability distribution of an energy measurement on $ \pmb \rho_6 $ [cf. equation~(\ref{eq:probabilities})]; namely as
\begin{eqnarray}
\mathcal{F}_\omega(\pmb H) &= \sum_{x=0}^{2^{n-1}-1} \left[\frac{(\partial_\omega p_{0,h(x)})^2}{p_{0,h(x)}}+\frac{(\partial_\omega p_{1,h(x)})^2}{p_{1,h(x)}}\right] \nonumber\\
& = \sum_{m=0}^{n-1} {{n-1}\choose{m}} \left[ \frac{(\partial_\omega p_{0,m})^2}{p_{0,m}} + \frac{(\partial_\omega p_{1,m})^2}{p_{1,m}} \right].
\label{eq:cfi}
\end{eqnarray}
When evaluating these derivatives, one must bear in mind that $ R = \frac{\gamma_0}{\lambda \epsilon} $ does depend on $ \omega $, as $ \epsilon = \tanh{\big(\frac{\omega}{2T}\big)} $. However, in our model $ \mathcal{F}_\omega(\pmb H) $ may be well approximated by taking $ R $ and $ \epsilon $ as constants, in the limit $ R\lambda \ll 1 $. That is,
\begin{eqnarray}
\mathcal{F}_\omega(\pmb H) &\simeq \sum_{m=0}^{n-1}{{n-1}\choose{m}}\nonumber\\
&\times\frac{4(a_m+b_m)c_m^2(n-2m)^2 t^2 \sin^2[\zeta_2+(2m-n)(\zeta_1 + t\omega)]}{(a_m + b_m)^2 - 4 c_m^2 \cos^2{[\zeta_2 + (2m-n)(\zeta_1 + t \omega)]}}.
\label{eq:cfi_smallR}
\end{eqnarray}

For even $ n $, the measurement setting $ (\zeta_1,\zeta_2) = (\frac{\pi}{2}-\bar{\omega} t,\frac{\pi}{2}) $ maximises $ \mathcal{F}_\omega(\pmb H) $, while for odd $ n $, one needs to choose $ (\zeta_1,\zeta_2) = (\frac{\pi}{2}-\bar{\omega} t, 0) $. Note that $ \bar{\omega} $ should not be thought-of as a variable, but as the best available estimate of the atomic frequency at any given stage. As the knowledge about $ \omega $ is refined, the value of $ \bar{\omega} $ should be updated, and the measurement setting, adaptively modified. Although it may seem counter-intuitive, \textit{undoing} the precession $ \pazocal{U}_{\omega t}^{\otimes n} $ on all atoms after the free evolution, improves the sensitivity to small fluctuations of $ \omega $ around its average $ \bar{\omega} $ and thus, helps to reduce $ \delta\omega $.

\begin{figure*}
	\includegraphics[width=0.5\linewidth]{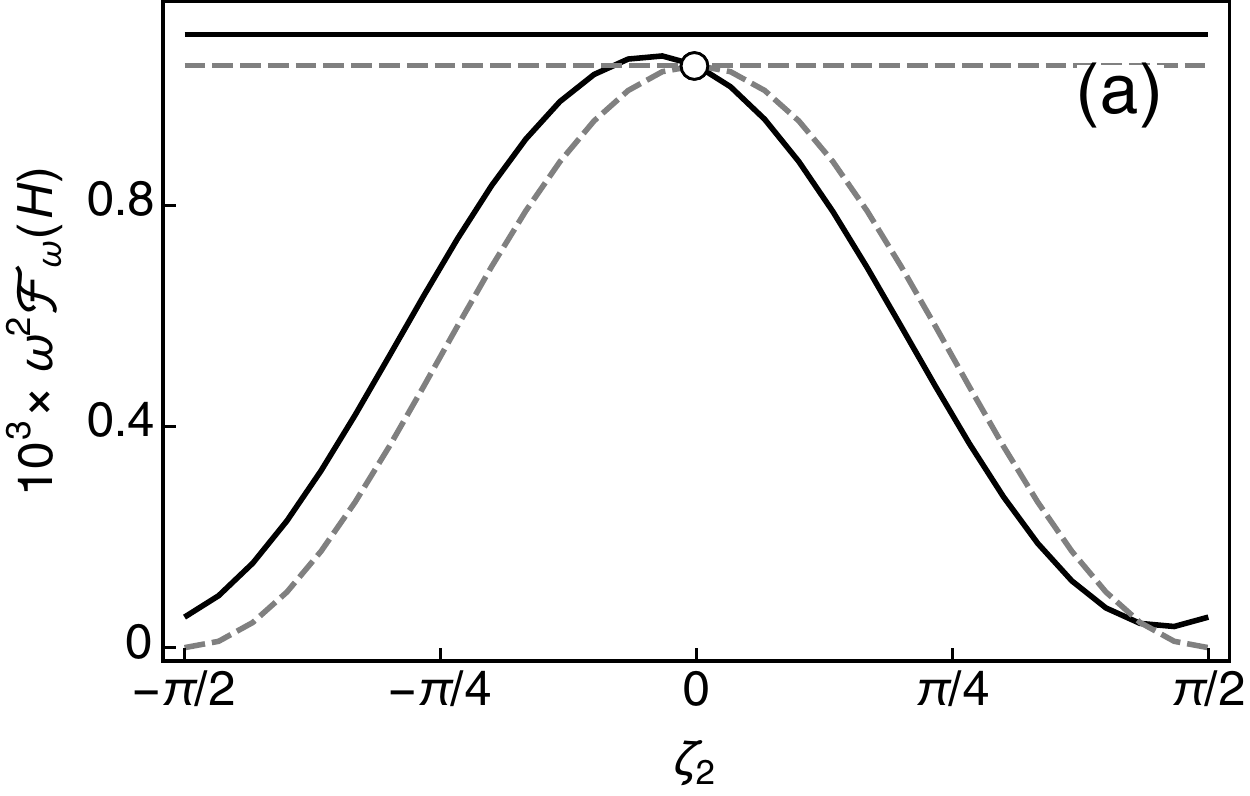}
	\includegraphics[width=0.49\linewidth]{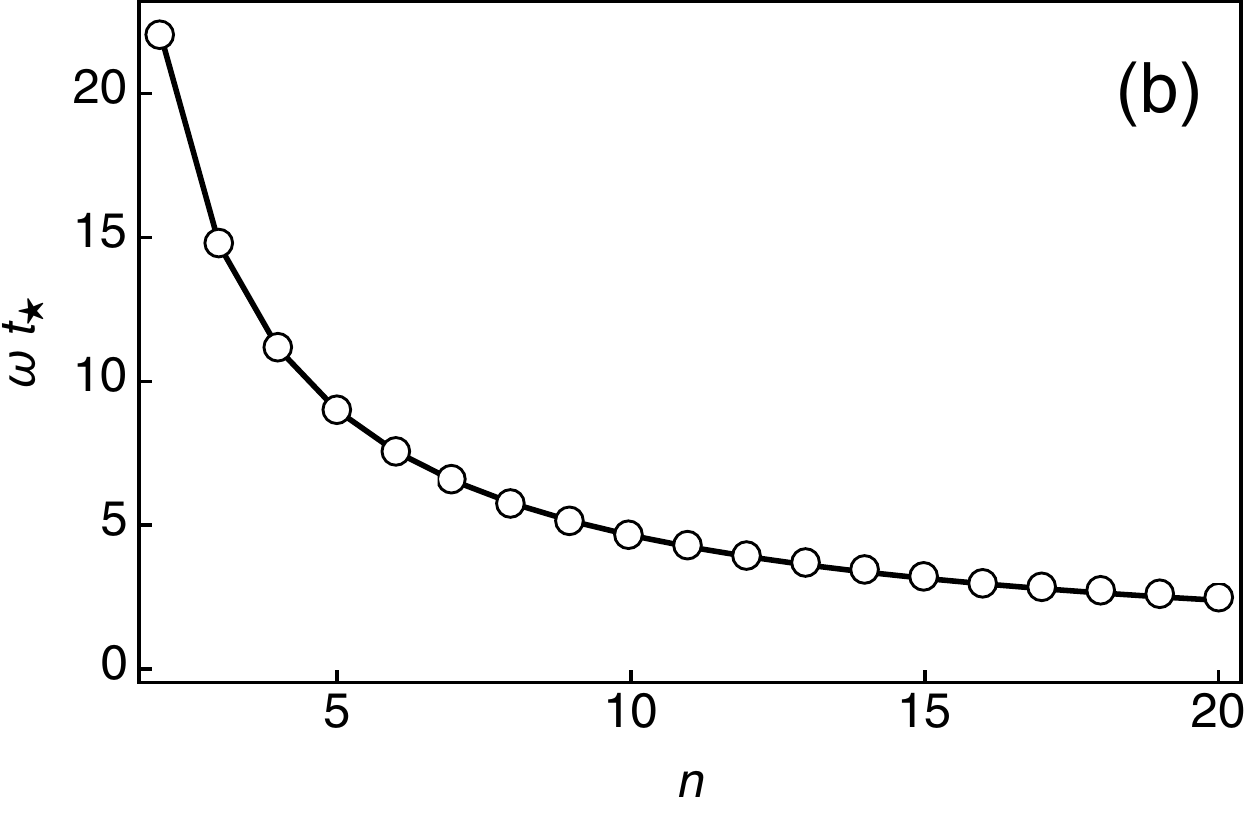}
	\caption{{\sf (a)} Approximate $ \mathcal{F}_\omega(\pmb H) $ for small $ R\lambda $, as in equation~(\ref{eq:cfi_smallR}), (dashed grey curve) and exact Fisher information (solid black curve), as compared with the approximate QFI of equation~(\ref{eq:qfi}) (dashed grey line) and the exact QFI (solid black line). The angle $ \zeta_1 $ is set to $ \zeta_1 = \frac{\pi}{2} - \bar{\omega} t $. Note the intersection of the curves at the nearly optimal measurement setting $ \zeta_2 = 0 $. {\sf (b)} Optimal interrogation time $ t_\star\sim n^{-1} $ as a function of the size of the probe $ n $. In both plots $ \omega = \bar{\omega} = 1 $, $ T = 200 $, $ \gamma_0 = 10^{-4} $, $ \lambda = 5 $ ($ R\lambda = 0.2 $), and $ t = 1 $. In {\sf (a)}, $ n=9 $.}
	\label{fig1}
\end{figure*}

\subsubsection{Optimality of the measurement scheme---\kern -0.8em} We now answer the question of whether another observable $ \pmb O \neq \pmb H $ may give a better frequency estimate by comparing $ \mathcal{F}_\omega(\pmb H) $ with the quantum Fisher information (QFI) $ F_\omega = \sup_{\pmb O} \mathcal{F}_\omega(\pmb O) $ \cite{paris2009quantum,kolodynski2014precision}. This can be computed from the state $ \pmb\rho_4 $ right after the free evolution stage or, equivalently, from $ \pmb\rho_5 $, as $ F_\omega $ is invariant under unitary transformations. The QFI is \cite{0253-6102-61-1-08}
\begin{equation}
F_\omega = 4\sum_{s,s'=\pm} \frac{\nu_x^s}{(\nu_x^s + \nu_x^{s'})^2} \vert \langle\Xi_x^s\vert\partial_\omega\pmb\rho_5\vert\Xi_x^{s'}\rangle \vert^2,
\label{eq:qfi_general}
\end{equation}
where $ \nu_x^{\pm} $ and $ \ket{\Xi_x^\pm} $ are the eigenvalues and eigenvectors of $ \pmb\rho_5 $. Specifically, these are
\begin{eqnarray}
\nu_x^{\pm} &= \frac12(a_x + b_x \pm \Delta_x) \nonumber \mbox{ and} \\
\ket{\Xi_x^\pm} &= \frac{(a_x-b_x\pm\Delta_x)\ket{0} + 2 c_x e^{i\omega t f(x)} \ket{1}}{\sqrt{4 c_x^2 + (a_x-b_x\pm\Delta_x)^2}}\otimes\ket{x},
\label{eq:eigensystem_rho5}
\end{eqnarray}
where $ \Delta_x \equiv \sqrt{(a_x-b_x)^2+ 4 c_x^2} $. Once again, we place ourselves in the limit of small $ R\lambda $, and find that $ \bra{\Xi_x^\pm}\partial_\omega\pmb\rho_5\ket{\Xi_x^\mp} = 0 $, and thus
\begin{equation}
F_\omega \simeq \sum_{m=0}^{n-1} {{n-1}\choose{m}}\frac{4(n-2m)^2 t^2 c_m^2}{a_m + b_m},
\label{eq:qfi}
\end{equation}
which exactly coincides with the maximum of equation~(\ref{eq:cfi_smallR}). Therefore, our proposed measurement setting is indeed optimal for $ R\lambda\ll 1 $. For arbitrary $ R\lambda $, however, $ F_\omega $ can be significantly larger than its limiting value (\ref{eq:qfi}). It may even be impossible to find a pair $ (\zeta_1,\zeta_2) $ so that $ \mathcal{F}_\omega(\pmb H) = F_\omega $. Nevertheless, the exact $ \mathcal{F}_\omega(\pmb H) $ \textit{always} coincides with (\ref{eq:qfi}) at $ \zeta_1 = \frac{\pi}{2} - \bar{\omega} t $ and $ \zeta_2 = \{ \frac{\pi}{2}, 0 \} $, even when this measurement setting is \textit{sub-optimal}. This point is illustrated in figure~\ref{fig1}(a).

\section{Results and discussion}\label{sec:results}

\begin{figure*}
	\includegraphics[width=0.44\linewidth]{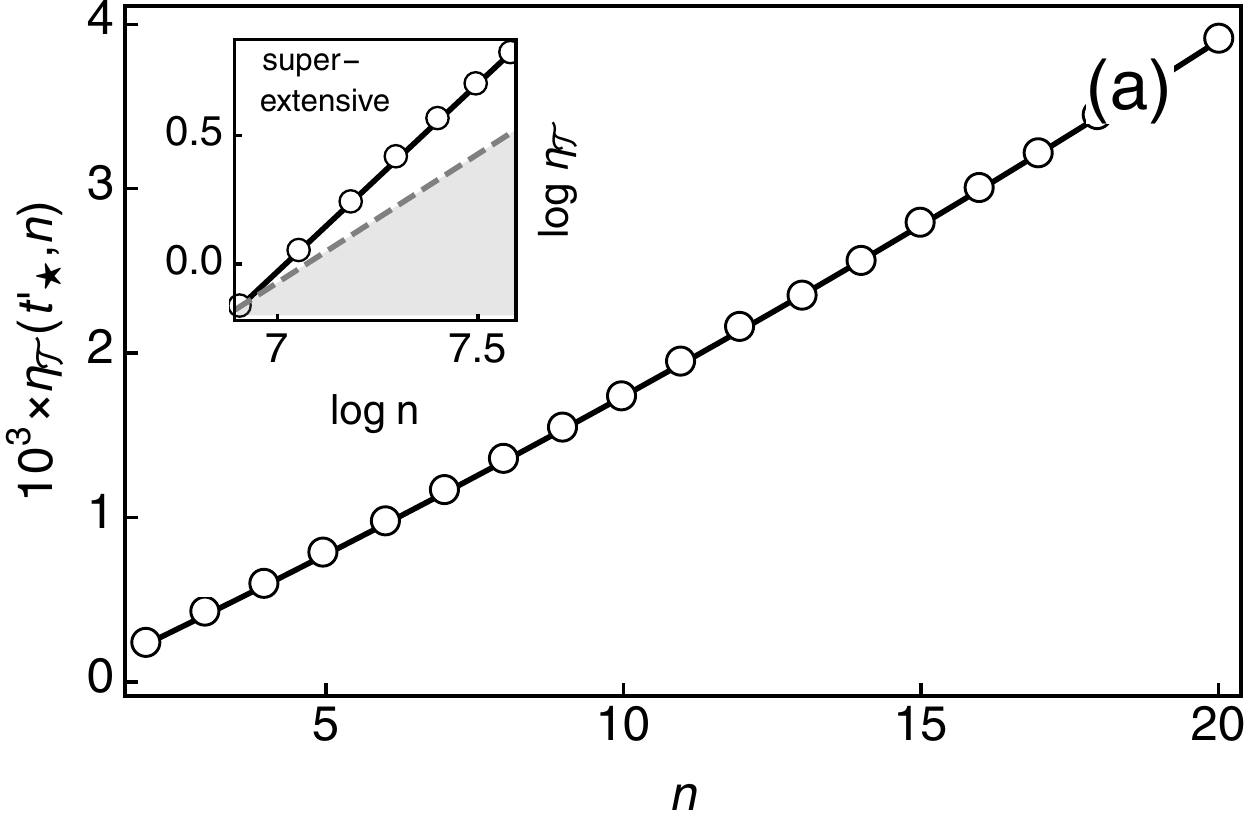}
	\includegraphics[width=0.26\linewidth]{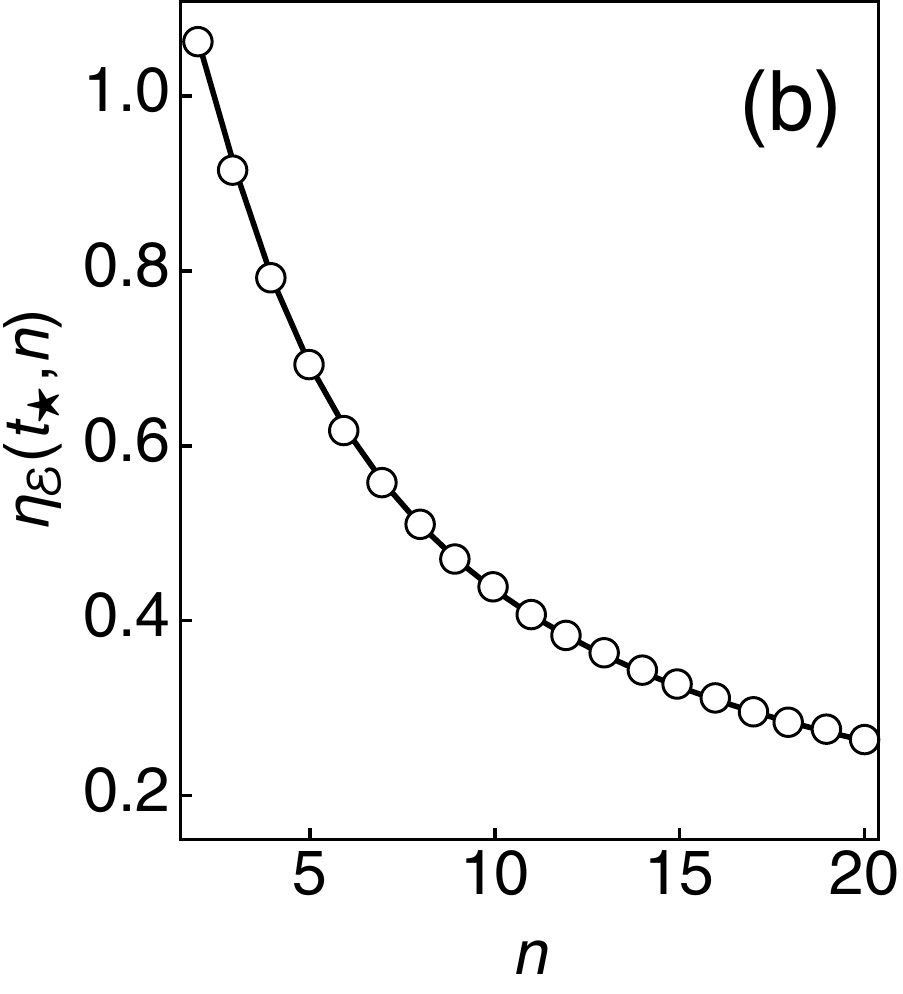}
	\includegraphics[width=0.27\linewidth]{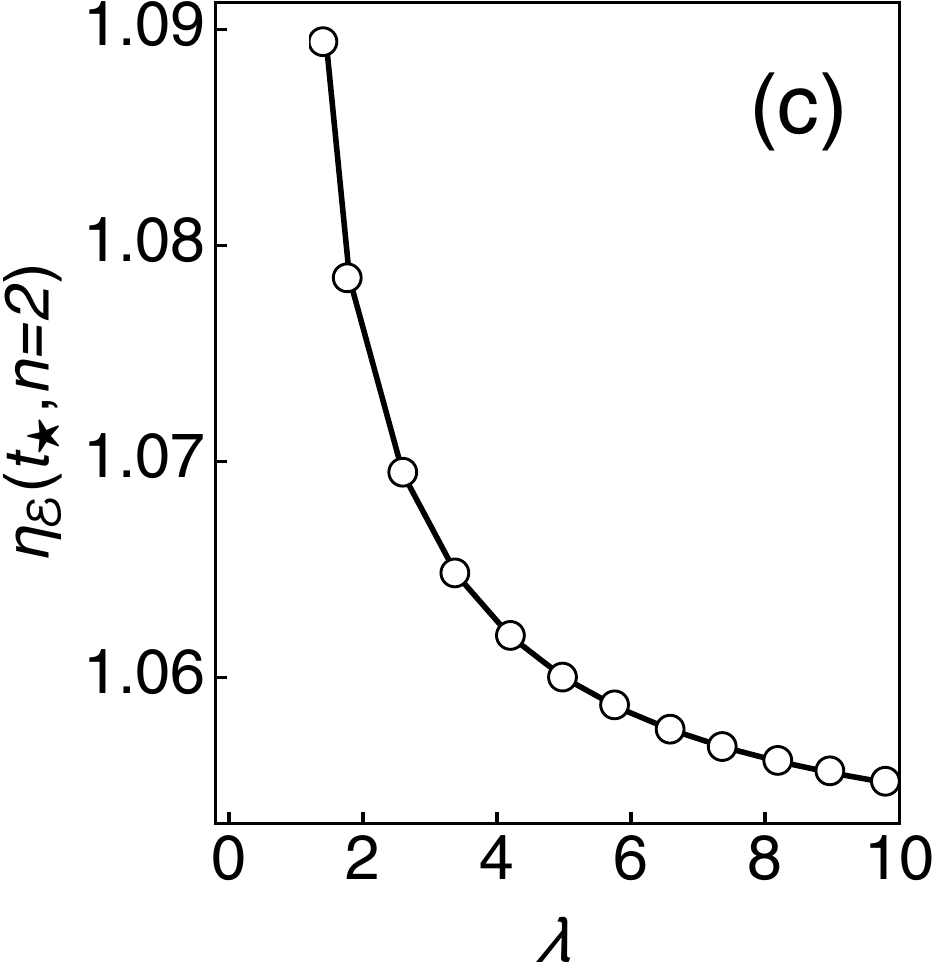}
	\caption{{\sf (a)} Efficiency $ \eta_{\pazocal{T}}(t'_\star,n) = F_\omega/t'_\star $ at the optimal interrogation time $ t'_\star $ as a function of the probe size $ n $, in the standard frequency-estimation scenario of limited time $ \pazocal{T} $. Note from the inset that, in spite of the fact that the probe is prepared in a \textit{mixed} GHZ-diagonal state, the efficiency grows super-extensively, as $ \tilde{\eta}(t_\star,n) \sim n^{3/2} $, which corresponds to Zeno scaling. {\sf (b)} Energy-efficiency $ \eta_{\pazocal{E}}(t_\star,n) = F_\omega/(\pazocal{E}_{init} + \pazocal{E}_{meas}) $ at the optimal interrogation time $ t_\star $ as a function of the probe size $ n $ for the same parameters as {\sf (a)}. In this case, one roughly has $ \eta_{\pazocal{E}}(t_\star,n)\sim n^{-1/3} $, i.e. from an energetic perspective, using large entangled probes yields no metrological advantage. In {\sf (c)}, we set $ n=2 $ and investigate how $ \eta_{\pazocal{E}} $ at $ t_\star $ decays as $ \lambda $ grows; that is, in our model, longer memory times yield more energy-efficient frequency estimation than purely Markovian dissipation. All parameters are the same as in figure~\ref{fig1}.}
	\label{fig2}
\end{figure*}

Recall that, in our scheme, the number of data points $ M $ that enters the inequality $ \delta\omega \geq 1/\sqrt{M \mathcal{F}_\omega(\pmb H)} $ is limited by the available energy $ \pazocal{E} $ as $ M = \pazocal{E}/(\pazocal{E}_{init} + \pazocal{E}_{meas}) $. We can thus define the energy efficiency
\begin{equation}
\eta_{\pazocal{E}}(t,n) \equiv \frac{F_\omega}{\pazocal{E}_{init}+\pazocal{E}_{meas}}.
\label{eq:efficiency}
\end{equation}
Note that we use $ \mathcal{F}_\omega(\pmb H) $ and $ F_\omega $ indistinctly since, for $ R\lambda \ll 1 $, the QFI becomes saturable with our optimal measurement prescriptions.

We will proceed to maximise $ \eta_{\pazocal{E}}(t,n) $ in two steps: First, for given $ n $, we shall find the optimal interrogation time $ t_\star $. Then, we will look at the scaling of $ \eta_{\pazocal{E}}(t_\star,n) $ with the probe size. From equations (\ref{eq:energy_preparation}), (\ref{eq:energy_rho_4}), (\ref{eq:energy_rho_6}), and (\ref{eq:qfi}), $ t_\star $ can be found numerically. As shown in figure~\ref{fig1}(b) it has a power-law-like dependence on the probe size $ \omega t_\star\propto n^{-c} $, where $ c \lesssim 1 $ (for $ R\lambda \ll 1  $).

Let us place ourselves in the standard scenario, in which the total time $ \pazocal{T} $ is the scarce resource to `economise' on. As usual, we shall work in the limit $ R\lambda \ll 1 $ and denote the corresponding optimal sampling time by $ t_{\star}' $, respectively. In figure~\ref{fig2}(a) we illustrate that $ \eta_{\pazocal{T}}(t_\star',n) $ can scale super-extensively under our time-inhomogeneous dissipative dynamics---even if we start from (mixed) thermal probes. Specifically, we recover the Zeno scaling $ (\delta\omega)^2 \sim 1/n^{3/2} $ \cite{PhysRevA.92.010102, PhysRevLett.109.233601}.

What figure~\ref{fig2}(a) suggests is that, if a large number $ N $ of two-level atoms were available, it would be sensible to batch them together in an entangled GHZ-diagonal state and partition the available running time $ \pazocal{T} $ into prepare-and-measure segments of length $ t'_\star $---the larger the probe, the better the resulting estimate.

In contrast, figure~\ref{fig2}(b) tells a completely different story: When adopting an entangled GHZ-diagonal preparation, the efficiency $ \eta_{\pazocal{E}}(t_\star,n) $ \textit{decreases} rapidly as the probe is scaled up in size (in this case $ \eta_{\pazocal{E}}(t_\star,n)\sim n^{-1/3} $, although the exponent is non-universal). This is so because, while $ (\pazocal{E}_{init} + \pazocal{E}_{meas}) \sim n $, the QFI exhibits a slower power-law-like growth. Hence, if there was a cap on the total available energy $ \pazocal{E} $, one could produce a more accurate frequency estimate by manipulating the uncorrelated atoms \textit{locally} rather than attempting to build such an `expensive' entangled state. Our numerics show that this qualitative behaviour persists even if we move away from the regime of $ R\lambda \ll 1 $ and search for the measurement setting $ (\zeta_1,\zeta_2) $ \textit{and} interrogation time $ t_\star $ which jointly maximise $ \eta_{\pazocal{E}}(t,\zeta_1,\zeta_2,n) = \mathcal{F}_\omega(\pmb H)/[\pazocal{E}_{init} + \pazocal{E}_{meas}(\zeta_1,\zeta_2) ] $.

Another natural question to ask in this setting is whether the environmental memory time plays any role in the energy efficiency of frequency estimation. In figure \ref{fig2}(c) we illustrate how $ \eta_{\pazocal{E}}(t_\star,n) $ decays with $ \lambda $ at any given $ n $. Recall from equation~(\ref{eq:phenomenological_master}) that increasing $ \lambda $ corresponds to reducing the bath memory time, thus making the dissipation `more Markovian'. Our setting thus showcases how memory effects in the dissipative dynamics can improve the performance of a specific parameter-estimation task. Elucidating whether memory effects play an instrumental role in energy-efficient frequency estimation requires a more general analysis that we defer for future work.

\section{Conclusions}\label{sec:conclusions}

We have studied the problem of noisy frequency estimation when the total available energy $ \pazocal{E} $ is limited. In each round of our estimation protocol, an ensemble of $ n $ initially thermal two-level atoms is brought into a GHZ-diagonal form by means of a simple sequence of qubit gates. We quantified the \textit{energetic cost} of the preparation stage $ \pazocal{E}_{init} $ by looking at the ensuing increase in the average energy of the probe.

The system is then allowed to evolve freely under the effect of environmental noise. This is modelled by a phenomenological master equation with built-in memory effects, which gives rise to phase-covariant free dissipative dynamics.

After further qubit operations, an energy measurement is eventually performed on the probe. We showed that, in a suitable range of parameters, these operations can be chosen so as to globally minimise the statistical uncertainty of the final frequency estimate. We also provided the corresponding optimal measurement prescription explicitly. The cost associated with the (pre-)measurement stage $ \pazocal{E}_{meas} $ can also be readily calculated from the change in the average energy of the probe, thus allowing for a comprehensive energetic bookkeeping in each round of the protocol.

We introduced the notion of energy efficiency of the estimation $ \eta_{\pazocal{E}}=\mathcal{F}_\omega(\pmb H)/(\pazocal{E}_{init} + \pazocal{E}_{meas}) $ as a means to assess the overall performance of the estimation protocol when there is a cap on the total energy $ \pazocal{E} $. We further found the optimal free-evolution time $ t_\star $ maximising $ \eta_{\pazocal{E}}(t_\star,n) $, and noticed that preparing larger probes in entangled GHZ-diagonal states is \textit{always detrimental} for the energy efficiency of frequency estimation.

In the standard scenario, one assumes that the most restrictive constraint is instead the limited running time $ \pazocal{T} $ of the estimation protocol and resorts to the figure of merit $ \eta_{\pazocal{T}} = \mathcal{F}_\omega(\pmb H)/t $. This grows monotonically with $ n $ when optimised over the free evolution time of the probe, thus suggesting that large multipartite entangled probes are, in principle, better. This is so because a figure of merit like $ \eta_{\pazocal{T}} $ fails to capture how `difficult' or `costly' it may be to prepare those states in practice. Incorporating the energetic dimension to the performance assessment through our $ \eta_{\pazocal{E}} $ may be the simplest way to quantitatively account for this `difficultness'.

It is true that tracking the average energy changes of the probe may be a crude way of capturing the actual limitations in force in real metrological setups. Likewise, in many situations, the total time $ \pazocal{T} $ might indeed place the most stringent limitation on the achievable precision, thus rendering other considerations irrelevant. Our observation merely highlights the importance of formulating quantifiers of the metrological efficiency that faithfully capture \textit{all} the relevant constraints in place in each specific scenario.

We also showed that, at any probe size, $ \eta_{\pazocal{E}}(t_\star,n) $ decays monotonically with the inverse bath memory time $ \lambda $, hence suggesting that large bath correlation times might be a resource for energy-efficient frequency estimation. This point certainly deserves a deeper and more general investigation.

Our intended take-home message is that \textit{different assessments of resources lead to different notions of optimality}. Hence, in order to produce practically useful metrological bounds, the stress should be placed on searching for those figures of merit capable of capturing the most stringent limitations at work in each experimental setup.

To conclude, it is important to remark that we did not optimise our energy efficiency over the initial state of the probe but rather, adopted the GHZ-diagonal preparation as a working assumption. The question of whether or not other forms of multipartite sharing of correlations could give rise to a more energetically favourable scaling remains open and certainly deserves further investigation. 

\subsubsection*{Acknowledgements---\kern -0.5em} We are thankful to A. del Campo, K. V. Hovhannisyan, J. Ko\l{}ody\'{n}ski, R. Kosloff, K. Macieszczak, M. Mehboudi, J. Oppenheim, R. Nichols,  N. A. Rodriguez-Briones, A. Smirne, T. Tufarelli, and R. Uzdin for helpful comments. We gratefully acknowledge funding from the Royal Society under the International Exchanges Programme (Grant No.~IE150570), the European Research Council under the StG GQCOP (Grant No.~637352), the Foundational Questions Institute (fqxi.org) under the Physics of the Observer Programme (Grant No.~FQXi-RFP-1601), and the COST Action MP1209: ``Thermodynamics in the quantum regime''.
	
\section*{References}

\providecommand{\newblock}{}

\end{document}